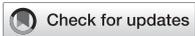

# Bridging high resolution sub-cellular imaging with physiologically relevant engineered tissues

Yasaman Kargar Gaz Kooh[1] and Nathaniel Huebsch[2]*

[1]Institute of Material Science and Engineering, Washington University in St. Louis, St. Louis, MO, United States, [2]Department of Biomedical Engineering, Washington University in St. Louis, St. Louis, MO, United States

While high-resolution microscopic techniques are crucial for studying cellular structures in cell biology, obtaining such images from thick 3D engineered tissues remains challenging. In this review, we explore advancements in fluorescence microscopy, alongside the use of various fluorescent probes and material processing techniques to address these challenges. We navigate through the diverse array of imaging options available in tissue engineering field, from wide field to super-resolution microscopy, so researchers can make more informed decisions based on the specific tissue and cellular structures of interest. Finally, we provide some recent examples of how traditional limitations on obtaining high-resolution images on sub-cellular architecture within 3D tissues have been overcome by combining imaging advancements with innovative tissue engineering approaches.

# 1 Introduction

High-resolution sub-cellular imaging has greatly advanced the cell biology field, making it possible to visualize organelles and other critical sub-cellular structures and molecular assemblies (D'Este et al., 2024; Lučić et al., 2013). For example, the integration of high-resolution techniques like multiphoton microscopy (MPM) has shown promising diagnostic accuracy in assessing pediatric tissues and tumors (Goedeke et al., 2019). This has enhanced our understanding of normal developmental processes, but has also allowed assessment of cellular changes during disease development and progression (Mojahed et al., 2022; Hayes and Melrose, 2023).

The field of tissue engineering has made significant advancements in replicating the complex structure and function of biological tissues. Using these engineered tissues to mimic 3D complex human tissues has allowed researchers to obtain a more accurate understanding of disease mechanisms, whilst also enabling screening of pharmaceuticals to assess drug efficacy and toxicity. In the cardiovascular field, 3D engineered heart tissues have allowed researchers to study cell-cell and cell-ECM (extracellular matrix) interactions that cannot be studied in 2D monocultures (Simmons et al., 2024; Van Spreeuwel et al., 2014; Thavandiran et al., 2013; Michas et al., 2022; Ma et al., 2023). For example, in a recent study from our team, we found that induced pluripotent stem cell (iPSC) derived cardiomyocytes with a hypertrophic cardiomyopathy (HCM) associated genotype

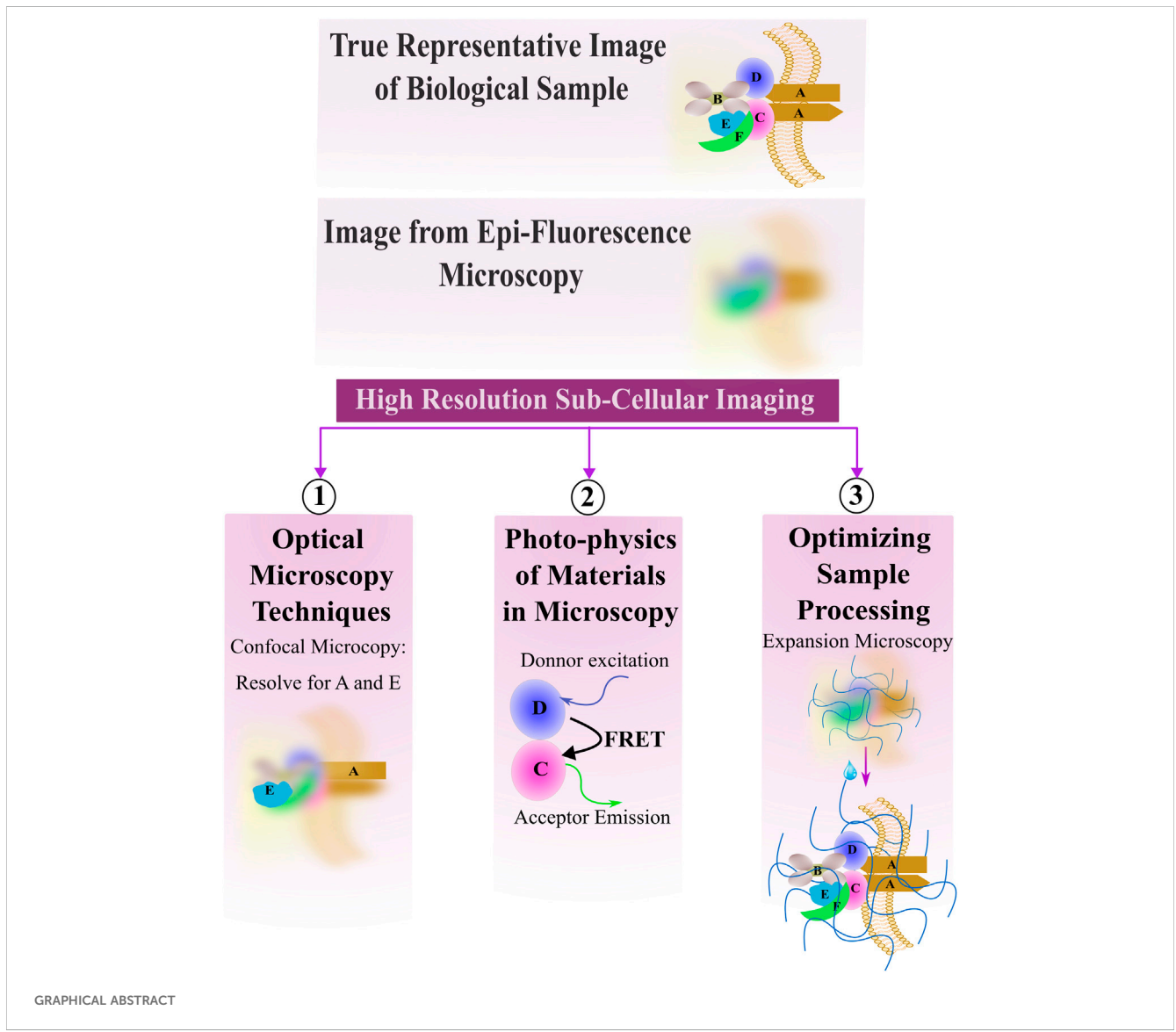

GRAPHICAL ABSTRACT

exhibited structural, pharmacologic and physiologic hallmarks of HCM in 3D engineered tissues that we were unable to appreciate in a traditional 2D culture (Guo et al., 2024). Analogously, in studying arrhythmogenic cardiomyopathy (ACM), a disease that has been linked to contractile deficits *in vivo*, (Chen et al., 2024) observed that isolated iPSC-cardiomyocytes with an ACM-linked genotype were largely similar to (if not *more* contractile than) isogenic controls. However, growing the iPSC-cardiomyocytes with an ACM genotype in 3D engineered tissues unveiled deficits in contractility that were attributed to weaker cell-cell mechanical coupling (Zhang et al., 2021).

In cancer research, 3D tumor models provide a realistic environment for studying tumor progression and testing anticancer drugs (Kang et al., 2020). Drug discovery and toxicology also benefit from 3D engineered tissues, since they offer more accurate models for evaluating drug efficiency and safety compared to traditional 2D cell cultures (Khetani and Bhatia, 2008; Khalil et al., 2020; Simmons et al., 2022). For example, a 3D tumor spheroid model incorporating stromal fibroblasts has been developed to better mimic the *in vivo* tumor microenvironment. This model not only enables the study of tumor-stroma interactions and the role of cancer-associated fibroblasts (CAF) in cancer progression but also facilitates drug discovery (Shao et al., 2020).

Besides their translational role in disease modeling and drug development, engineered tissues have also allowed scientists to study fundamental questions in developmental biology (including tissue morphogenesis and cell differentiation; (Zuncheddu et al., 2021; Pien et al., 2023)), and biophysics [including the mechanical properties and behaviors of various tissues under different physiological conditions; (Guo et al., 2021a; Li et al., 2023)]. The development of organoid technology, in which "organ-like" structures develop from progenitor and/or stem cells *in vitro* to recapitulate key aspects of normal tissue development, has allowed scientists to study "organs in a dish" that replicate the phenotypic and functional properties of real tissue environments (Moysidou et al., 2021; Bose et al., 2019; Bhatia and Ingber, 2014). Importantly, organoids allow *in vitro* differentiation of cell types (notably, certain endodermal cell populations) that have historically proven challenging to derive through directed differentiation methods (Vunjak-Novakovic et al., 2021; Hofer and Lutolf, 2021).

Given that cells exhibit more "*in vivo*-like" signaling, pharmacology and behavior within 3D engineered tissues, organoids and other complex *in vitro* assembled structures compared to what they would exhibit in more "standard" 2D monoculture environments, a longstanding question for bioengineers and cell biologists is "how are cells sensing their environment?." Forces that cells experience are dramatically different in 3D tissues as opposed to standard 2D culture, and indeed, recent studies by our team and others, suggest that mechanical stretch affects electrophysiology of engineered heart muscle (Simmons et al., 2024; Guo and Huebsch, 2020; Wang C. et al., 2023; Schuftan et al., 2024; Boudou et al., 2012; DePalma et al., 2023; Abilez et al., 2018; Tulloch et al., 2011; Huebsch, 2019; Guo et al., 2021b). This may be linked to either improved expression and/or intracellular transport of ion channels (Simmons et al., 2024; Marchal et al., 2021). In a broader context, Chaudhuri et al. provided a comprehensive review of the complex mechanical properties of tissue and ECMs including viscoelasticity, viscoplasticity, and nonlinear elasticity, and discussed how these properties affect cellular behaviors (Chaudhuri et al., 2020). Adding to this, Gong et al. found that maximum cell spreading occurs at an optimal viscosity level on soft substrates for low ECM rigidity, where the substrate relaxation time falls between clutch binding and lifetime timescales, enhancing cell-ECM adhesion, while on stiff substrates, viscosity has no effect due to clutch saturation, which provides insights for designing biomaterials to optimize cell adhesion and mechano-sensing (Gong et al., 2018). Moreover, in 3D cultures, cancer cells exhibit enhanced drug resistance mechanisms compared to 2D cultures, this is attributed to the more realistic cell-cell and cell-matrix interactions in 3D environments, which affect gene expression profiles and signaling pathways related to drug metabolism and resistance. Hypoxic conditions and nutrient gradients present in 3D tumor spheroids activate hypoxia-inducible factors (HIFs), leading to the upregulation of drug resistance genes and proteins (Bloise et al., 2024; Atat et al., 2022). Additionally, cancer cells in 3D environments use invadopodia to sense and dynamically respond to the mechanical properties of the ECM, such as its stiffness and plasticity. Through chemo-mechanical feedback, invadopodia can adjust their growth patterns, enabling cancer cells to invade tissues more effectively through chemo-mechanical signaling feedback loops (Gong et al., 2021). Therefore, understanding the molecular mechanisms cells use to sense their 3D environment will allow bioengineers to develop better *in vitro* model systems and may provide molecular targets that can be directly exploited for therapy.

A challenge in unveiling these sensing mechanisms is that such studies will require detailed analysis of subcellular architecture. This presents a fundamental technical challenge: these environmental changes that provoke a more "*in vivo*-like phenotype" requires putting cells into 3D, whereas traditionally, detailed analysis of subcellular architecture has required growing cells in 2D environments (e.g., culture on glass coverslips). While artificial, this culture approach places cells very close (typically, within 150 μm) to microscope objectives, allowing researchers to use high-resolution optics with high numerical-aperture immersion media (e.g., oil) and in turn facilitating robust analysis of sub-cellular architecture (Zuncheddu et al., 2021; Pien et al., 2023; Berry et al., 2021; Zhao et al., 2022; O'Connor et al., 2022; Tan et al., 2004).

Advancements in optics, coupled with advancements in the fluorescent probes used for imaging, offer promising means to combine high resolution subcellular imaging with physiologically relevant engineered tissues and organoids (Li et al., 2019; Dekkers et al., 2019; Hilzenrat et al., 2022; Wang S. et al., 2023). To achieve a high-resolution image, it is essential to balance spatial and temporal resolution, maintain an optimal signal-to-noise ratio, and avoid photobleaching (Figure 1). Here, we first review basic physical principles underlying fluorescence imaging. We then discuss promising approaches that tissue engineers have taken to enable high-resolution, sub-cellular imaging of 3D tissues, and provide examples of some of the key insights gained from these analyses. The lateral, axial, and temporal resolutions of these methods, along with their imaging depths, are detailed in Table 1, which also includes schematics of the focal plane for each technique. We aim to simplify the decision-making process for researchers choosing the most optimal imaging approach by presenting simplified guidance on choosing methods based on specific research needs. Excellent reviews of imaging methods, which delve more deeply into the physics underlying the various methods we discuss, are available elsewhere (Huang et al., 2009; Winter and Shroff, 2014; Leung and Chou, 2011). Another key point of our review is that while most high-resolution imaging methods focus on 2D samples or sectioned 3D tissue, we specifically highlight methods of whole-mount staining and optical sectioning for 3D tissues, which allows for better imaging of intact 3D structures, an area that remains underrepresented in the current literature.

# 2 Fundamentals of fluorescence microscopy: optical sectioning

Fluorescence microscopy is a cornerstone technique in biological sciences, enabling the visualization of cellular structures and functions by exploiting the principles of fluorescence (Masters, 2014). The underlying mechanism involves the excitation of fluorophores, molecules that can re-emit light upon excitation that is specifically attached to target molecules within the sample, such as proteins or nucleic acids. When exposed to light of a specific wavelength (typically from a UV source like a mercury arc lamp), these fluorophores absorb photons, exciting electrons to a higher energy state. The electrons, inherently unstable at this level, quickly return to their ground state, releasing photons with a longer wavelength and lower energy than the absorbed light (Sanderson et al., 2014). The change in wavelength between the photons that excite fluorescence and the photons emitted is called the Stokes shift (Peng et al., 2005). The Stokes shift occurs because the emitted light is of lower energy due to losses during the nonradiative transitions between excited and ground states, a process that Jablonski energy diagrams explains (Figure 2A) (Yao et al., 2014; Berezin and Achilefu, 2010). Detailed understanding of the physics of decay of the higher-energy states has allowed researchers to control Stokes shift and exploit non-radiative methods of energy decay (e.g., Förster Resonance Energy Transfer, FRET) for analyzing sub-cellular architecture (detailed in Section 3).

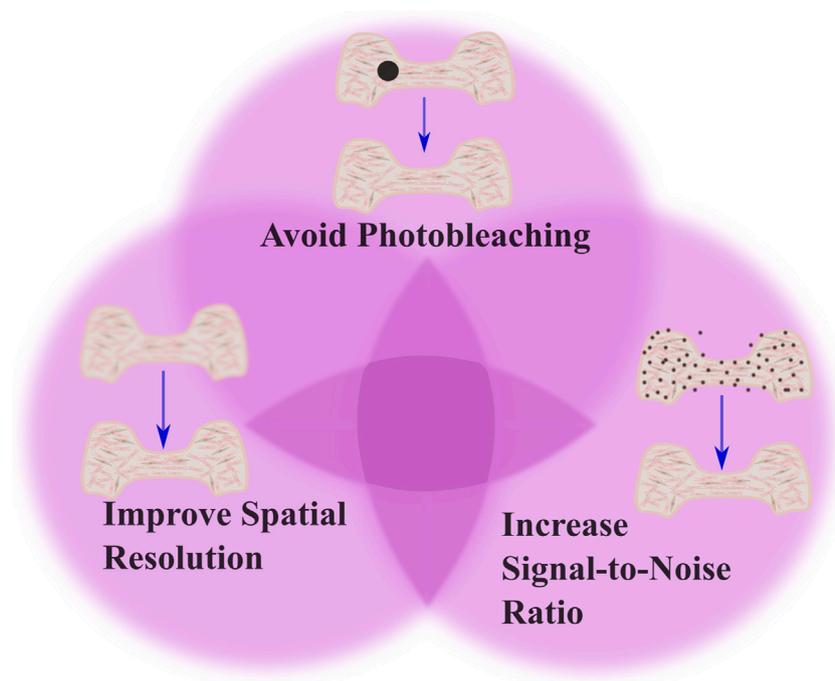

FIGURE 1
Tradeoffs in imaging techniques. The figure illustrates the balance between three critical factors in imaging: avoiding photobleaching, optimizing spatial resolution, and increasing signal-to-noise ratio (SNR). The top part of venn diagram highlights the importance of minimizing photobleaching, shown by the black dot indicating potential damage to the tissue from excessive illumination. The left part of the venn diagram emphasizes optimizing spatial resolution to capture detailed cellular processes. The right part of the venn diagram demonstrates the need to increase SNR, which improves image clarity by reducing noise. This is shown by the presence of fewer noise artifacts in the tissue structure. To achieve optimal imaging, there should be a balance among these factors to produce clear, accurate images while avoiding phototoxicity and photobleaching effects.

In fluorescence microscopy, optical filters and epi-illumination techniques are used to enhance quality of the images. An excitation filter selects the optimal wavelength to excite the fluorophore. A barrier filter blocks excitation wavelengths while passing longer emission wavelengths, and a dichroic mirror directs excitation light to the sample and emitted fluorescence to the detector. This allows the objective lens to both illuminate the specimen with the excitation light and to collect the emitted fluorescence (Figure 3A) (Lichtman and Conchello, 2005; Combs and Shroff, 2017). In diffraction limited (e.g., not super-resolution) optical systems, the minimum distance $d$ between objects that can be resolved is related to the wavelength of light ($\lambda$) and the numerical aperture ($N.A.$) is dictated by the equation

$$d = \frac{\lambda}{2N.A.}$$

Within the visible wavelength range, this sets a diffraction limit near 250–300 nm (Thorne and Blandford, 2017). This is significantly smaller than a mammalian cell (10 s of μm) but large compared to proteins (10 s of nm). Lower wavelength, higher energy photons (e.g., x-rays) yield a smaller diffraction limit that can allow resolving protein-protein interactions, but this often limits the ability to image live samples.

Atop the diffraction limit, epifluorescence microscopy methods face additional limitations in resolution and contrast due to the full-sample illumination strategy (Bates et al., 2008). While in theory the resolution limit of epifluorescence imaging is about 250 nm laterally and 500 nm axially (Lidke and Lidke, 2012), in practice it is often challenging to separate objects closer than 1 μm. The Point Spread Function (PSF) describes how light spreads as it passes through the microscope optics (Cole et al., 2011; Pankajakshan et al., 2009). In epifluorescence, there is high background intensity from out-of-focus fluorescence, limiting the clarity of the images (Figure 2B) (Huang et al., 2009; Sung et al., 2002; Huang et al., 2008a). These challenges have spurred the development of advanced microscopy techniques that enhance image quality by manipulating the light path and minimizing unwanted fluorescence (Heilemann, 2010; Hell, 2003; Hell, 2009; Helmchen and Denk, 2002). Such improvements not only reduce photobleaching effects, but also increase the throughput and specificity of fluorescence imaging systems, which allows for more detailed and quantitative studies of live cells and tissue (White, 2005; Ishikawa-Ankerhold et al., 2012). As researchers continue to refine these methods, the ability to visualize biological processes with greater precision enhances our understanding of complex cellular mechanisms within biological samples such as 3D engineered tissues.

## 2.1 Laser scanning confocal microscopy

Confocal microscopy was developed to enhance the optical resolution and contrast of images compared to what could be achieved with epifluorescence microscopy (Jonkman et al., 2020; Nwaneshiudu et al., 2012). In confocal microscopy, a laser beam is

TABLE 1 Comparison of key features of optical microscopy techniques.

| Technique | Lateral resolution | Axial resolution | Imaging depth | Typical speed* | Schematic of the focal plane | References |
|---|---|---|---|---|---|---|
| Epifluorescence Microcopy | >250 nm | >500 nm | 30 μm | 100 f.p.s | 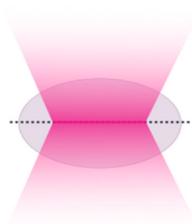 | Lidke and Lidke (2012) |
| Laser Scanning Confocal Microscopy | >200 nm | >500 nm | 100–150 μm | 10–30 f.p.s | 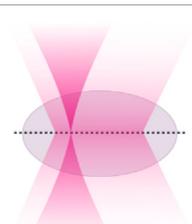 | Huebner et al. (2007), Fouquet et al. (2015), Choi et al. (2013) |
| Spinning Disc Confocal Microscopy | >200 nm | >500 nm | 150 μm | 40–400 f.p.s | 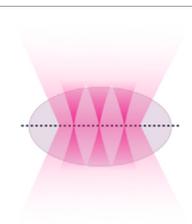 | Oreopoulos et al. (2014), Nakano (2002), Takahara et al. (2010) |
| Two Photon Microscopy | 300 nm | >500 nm | 500–1,000 μm | Varies with scanner type (1–30 f.p.s) | 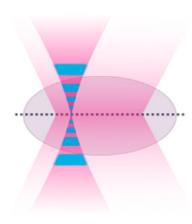 | Olivieri et al. (2014), Doi et al., 2018; Pawley (2006) |
| Three Photon Microscopy | 200 nm | 500 nm | >1,400 μm | Varies with scanner type (1–30 f.p.s) | | König et al. (2000), Streich et al. (2021) |
| Light Sheet Fluorescence Microscopy | 200 nm | 1,000 nm | >1 cm | 100 f.p.s | 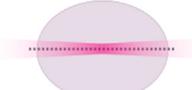 | Santi, 2011; Santi et al. (2009), Keller et al. (2008), Olarte et al. (2018), Chang et al. (2017) |
| Stimulated Emission Depletion (STED) microscopy | 20–50 nm | 30–50 nm | 120 μm | 30 f.p.s | 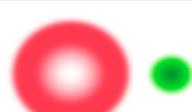 | Revelo et al. (2015), Wildanger et al. (2009), Wildanger et al. (2008), Blom and Brismar (2014), Tam and Merino (2015) |
| Stochastic Optical Reconstruction Microscopy (STORM) | 20–30 nm | 50–60 nm | 10 μm | Varies with the switching kinetics of the fluorophore (~1 f.p.s) | 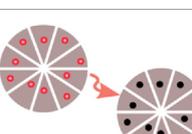 | Huang et al. (2008a), Xu et al. (2017), Vovard et al. (2024), Inal et al. (2021) |

*Note: provided speeds are for "full-frame" imaging, not subsampling, which is commonly used to achieve higher frame rates.

focused on a small point within the tissue and scans the specimen point-by-point, using a pinhole aperture in front of the detector to block out-of-focus light (Elliott, 2020). Thus allows only light from the focal plane to be detected (Figure 3B) (Paddock, 2014). This technique significantly reduces the noise from light scatter and out-of-focus blur (Wnek and Bowlin, 2008). The enhanced resolution and contrast in confocal microscopy are crucial for obtaining high-quality images of natural and engineered tissues, particularly in studies evaluating microstructure and cellular interactions within scaffolds (Dunkers et al., 2003). For example, interactions between collagen scaffolds and smooth muscle cells have been probes with confocal microscopy, allowing researchers to optimize scaffold design to enhance cell-ECM interactions (Sanderson et al., 2014; Amadori et al., 2006).

While superior to epifluorescence microscopy in terms of resolution limits, confocal microscopy faces its own set of challenges that limit practical applications. A primary issue is that the resolution often doesn't meet theoretical expectations due to limitations related to pinhole size and light intensity (Cox and Sheppard, 2004), since the resolution improvement requires an impractically small pinhole. Second, because the laser used for illumination must scan across pixels within the sample, rather than illuminate them all at once, laser scanning confocal microscopy is inherently slower than epifluorescence, limiting the ability to capture dynamics over larger scales (for example, across all cells within an engineered tissue). Assessing dynamic subcellular processes thus require researchers to balance the tradeoff between resolution and speed of image acquisition (Elliott, 2020). Third, photobleaching, wherein prolonged exposure to intense laser light degrades fluorescent dyes, is another significant challenge, which is worsened by the inherent need to expose regions within the sample to intense light for a prolonged time period in order to obtain detailed structural information (St. Croix et al., 2005). Having higher resolution can lead to unnecessary photobleaching without providing additional useful biological information (Bernas et al., 2004), and thus, researchers must carefully optimize imaging times. These challenges necessitate careful consideration of imaging parameters and objectives to ensure that the benefits of confocal microscopy outweigh drawbacks.

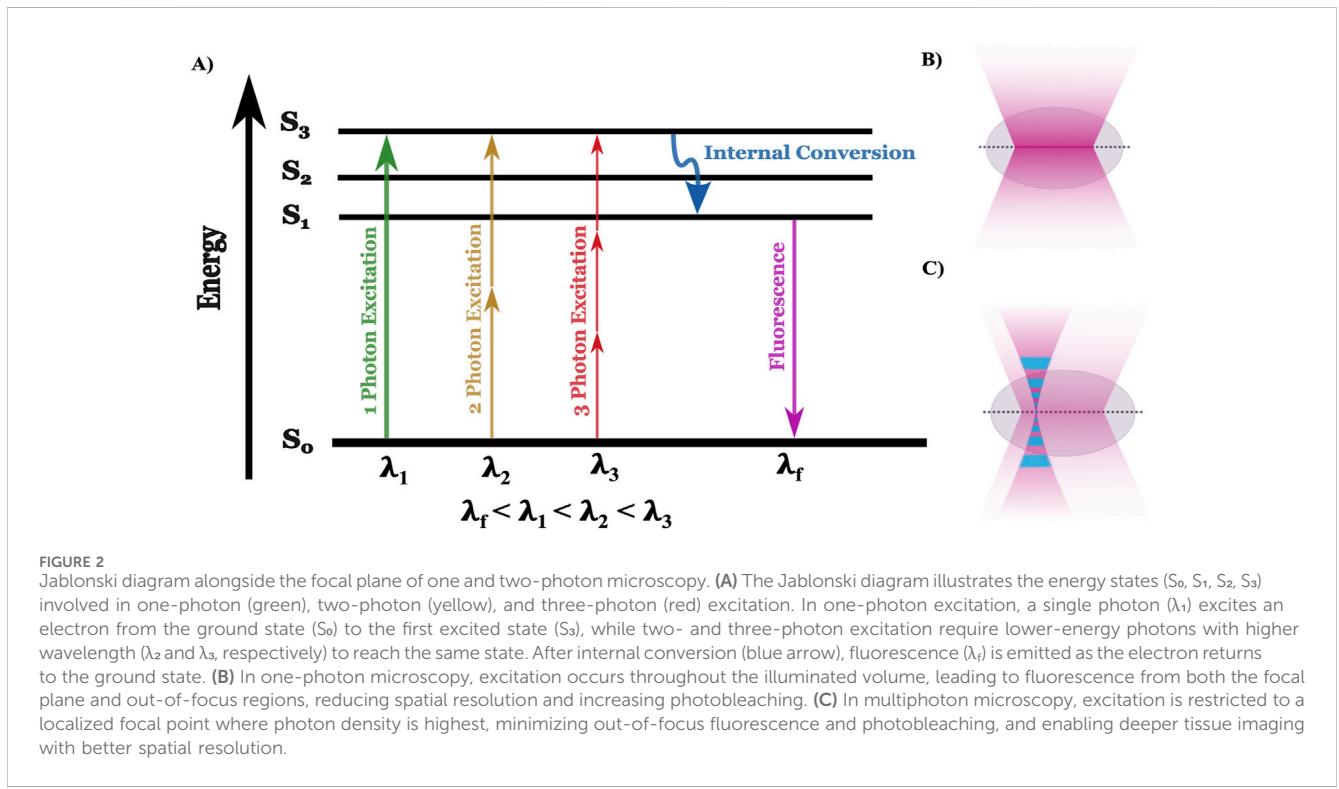

FIGURE 2
Jablonski diagram alongside the focal plane of one and two-photon microscopy. **(A)** The Jablonski diagram illustrates the energy states ($S_0$, $S_1$, $S_2$, $S_3$) involved in one-photon (green), two-photon (yellow), and three-photon (red) excitation. In one-photon excitation, a single photon ($\lambda_1$) excites an electron from the ground state ($S_0$) to the first excited state ($S_3$), while two- and three-photon excitation require lower-energy photons with higher wavelength ($\lambda_2$ and $\lambda_3$, respectively) to reach the same state. After internal conversion (blue arrow), fluorescence ($\lambda_f$) is emitted as the electron returns to the ground state. **(B)** In one-photon microscopy, excitation occurs throughout the illuminated volume, leading to fluorescence from both the focal plane and out-of-focus regions, reducing spatial resolution and increasing photobleaching. **(C)** In multiphoton microscopy, excitation is restricted to a localized focal point where photon density is highest, minimizing out-of-focus fluorescence and photobleaching, and enabling deeper tissue imaging with better spatial resolution.

## 2.2 Spinning disc confocal microscopy

To address the challenge of speeding up imaging acquisition and reducing photobleaching problems from confocal microscopy, spinning disk confocal microscopy was developed (Stehbens et al., 2012; Zimmermann and Brunner, 2006). Unlike traditional confocal microscopy, which employs a single pinhole to selectively capture light from the focal plane, spinning disc confocal microscopy utilizes a series of rapidly rotating pinhole-containing discs (Stehbens et al., 2012). These discs are positioned in alignment with the different focal planes of the specimen, and each pinhole targets a specific area on the sample (Figure 3C). As the discs spin, each pinhole scans across the specimen, rapidly acquiring images at multiple points simultaneously, which enhances light efficiency and reduces background interference compared to traditional confocal systems, which also significantly enhances imaging speed (Oreopoulos et al., 2014), effectively addressing previous limitations related to weak fluorescence signals (Zimmermann and Brunner, 2006). The increased imaging speed afforded by spinning disc confocal reduces photobleaching and phototoxicity by limiting the exposure of the sample to intense laser light. This technique also provides excellent optical sectioning capabilities, which allows imaging of 3D thick engineered tissues in more detailed by selectively collecting fluorescence signals from the focal plane while rejecting the out-of-focus light to produce high contrast and resolution (Oreopoulos et al., 2014; Wang et al., 2005).

Spinning disc microscopy is subjected to its own set of limitations. Most importantly, "crosstalk" of pinholes can generate a hazy background in images and impact the axial resolution negatively. The reason for this "crosstalk" is that the emitted light from one point of the specimen can enter multiple pinholes, especially when pinholes are closely spaced. This degrades image quality as it effectively convolves fluorescence emitted from different parts of the sample onto the same set of camera pixels (Oreopoulos et al., 2014). To solve this problem, systems have been developed that allow fine-tuning of pinhole size and spacing.

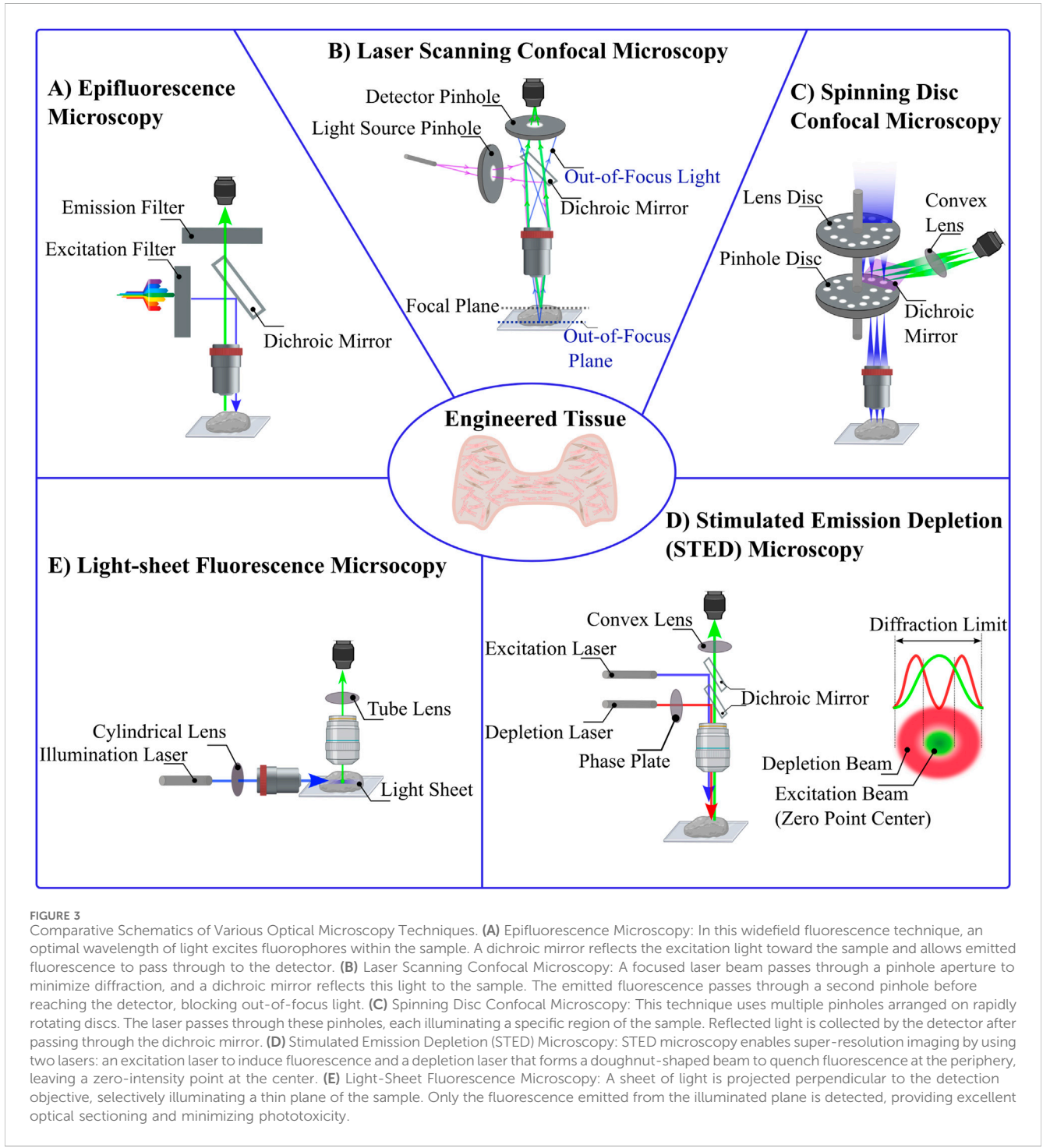

FIGURE 3
Comparative Schematics of Various Optical Microscopy Techniques. **(A)** Epifluorescence Microscopy: In this widefield fluorescence technique, an optimal wavelength of light excites fluorophores within the sample. A dichroic mirror reflects the excitation light toward the sample and allows emitted fluorescence to pass through to the detector. **(B)** Laser Scanning Confocal Microscopy: A focused laser beam passes through a pinhole aperture to minimize diffraction, and a dichroic mirror reflects this light to the sample. The emitted fluorescence passes through a second pinhole before reaching the detector, blocking out-of-focus light. **(C)** Spinning Disc Confocal Microscopy: This technique uses multiple pinholes arranged on rapidly rotating discs. The laser passes through these pinholes, each illuminating a specific region of the sample. Reflected light is collected by the detector after passing through the dichroic mirror. **(D)** Stimulated Emission Depletion (STED) Microscopy: STED microscopy enables super-resolution imaging by using two lasers: an excitation laser to induce fluorescence and a depletion laser that forms a doughnut-shaped beam to quench fluorescence at the periphery, leaving a zero-intensity point at the center. **(E)** Light-Sheet Fluorescence Microscopy: A sheet of light is projected perpendicular to the detection objective, selectively illuminating a thin plane of the sample. Only the fluorescence emitted from the illuminated plane is detected, providing excellent optical sectioning and minimizing phototoxicity.

## 2.3 Two-photon microscopy

In traditional confocal microscopy, high axial and lateral resolution are achieved using pinholes to effectively limit out-of-focus light from reaching the photo-detector. In contrast, multiphoton microscopy achieves high resolution by utilizing specialized excitation, exploiting nonlinear optical processes to allow visualization deep into tissue structures non-invasively with high spatial resolution (Schenke-Layland et al., 2006; Tsai et al., 2009; Neto et al., 2020; Diaspro et al., 2006). This method typically utilizes femtosecond laser pulses that minimizes heat damage while maintaining high photon flux to provide optical sectioning by exciting fluorescence only at the focal point of the microscope's objective which reduces photodamage and photobleaching (Schenke-Layland et al., 2005; Ustione and Piston, 2011; Levene et al., 2004; Zipfel et al., 2003). This is achieved through the simultaneous absorption of two or more lower-energy and higher-wavelength photons, a process that occurs only at very high photon densities (Ustione and Piston, 2011; Tauer, 2002). These densities are reached only at the focus due to the brief and intense laser pulses used. This selective excitation means that only the plane of focus emits fluorescence, creating a natural optical section without the need for physical sectioning of the sample, which produce sharp images with excellent contrast and minimal out-of-focus fluorescence (Dunn and Young, 2006).

Since it eliminates background very effectively, multiphoton microscopy allows for high-resolution imaging of thick, scattering biological specimens with minimal distortion, making it ideal for studying engineered tissues, live organisms, and dynamic cellular processes in their native environments (Tsai et al., 2009; Zipfel et al., 2003; Rubart, 2004). The selection of appropriate excitation wavelength, the number of the photons in excitation, and laser sources are important to maximize penetration depth and minimize photodamage (Cheng et al., 2014; Lefort, 2017; Kobat et al., 2009; Marsh et al., 2003; Jones et al., 2018; Sidani et al., 2006).

While it offers substantial advantages for deep tissue imaging, multiphoton microscopy is subject to unique challenges that merit careful consideration. The most important one is the potential for increased photobleaching and photodamage in the focal plane. Although, multiphoton excitation generally causes less photodamage compared to standard (one-photon) excitation, there is still potential for cellular effects when intensities at the focal point exceed certain thresholds like high-intensity NIR (700–1,000 nm) pulses (König, 2000). This damage may show as protein denaturation, oxidative stress, DNA damage, and reduced cell viability, potentially affecting the accuracy of biological studies. The risk of such forms of damage is particularly acute with shorter pulse widths (femtosecond pulses) and higher power settings that can lead to unintended three-photon excitation. In this case, three (rather than two) low energy photons combine to create an extremely high energy, low wavelength photon. Unintended 3-photon excitation is a rare event, but when it does occur, it can cause severe cellular damage, including DNA damage and plasma generation, leading to intense localized luminescence and potential tissue carbonization (Nadiarnykh et al., 2012). Another challenge with two-photon microscopy is that samples containing light-absorbing pigments like hemoglobin or melanin can experience significant heating effects, leading to physical cell damage. To mitigate these risks, lowering excitation rates and optimizing the focal volume through under-illumination of the back focal plane of the objective could be useful. These can help reduce photodamage while still capturing sufficient signal for effective imaging (Tauer, 2002; König, 2000). Finally, while offering superior depth penetration and reduced photodamage, multiphoton microscopy, like laser scanning confocal microscopy, may still fall behind other techniques in terms of temporal resolution and imaging speed, particularly when high-resolution imaging is required over large areas or volumes (König, 2000). This limitation primarily stem from the need to excite a very small region of interest (ROI) at a time in both techniques (Reddy et al., 2015).

## 2.4 Thin sheet (light sheet) fluorescence microscopy

Light-sheet fluorescence microscopy provides high-resolution, three-dimensional views of biological samples especially thick 3D engineered tissues, with minimal photodamage (Parthasarathy, 2018). These advantages make this method ideal for the three-dimensional imaging of live or fixed, small or large engineered tissues (Pampaloni et al., 2015; Stelzer et al., 2021). Light-sheet fluorescence microscopy, samples are illuminated by a plane of light oriented perpendicular to the detection lens, ensuring that only the focal plane is exposed to light at any given moment (Figure 3E). This significantly reduces light exposure, thus minimizing photodamage and photobleaching compared to conventional fluorescence microscopy methods (Santi, 2011; Zubkovs et al., 2018; Fei et al., 2016; Swoger et al., 2014). By limiting illumination to the thin section of the sample that is in focus, light-sheet fluorescence microscopy enhances both spatial and temporal resolution, enabling the detailed observation of dynamic biological processes in real-time (Fei et al., 2016; Mohan et al., 2014; Daetwyler and Huisken, 2016; Lim et al., 2014; Delgado-Rodriguez et al., 2022; You and McGorty, 2021). The use of orthogonal illumination and detection pathways allows for rapid imaging across large volumes, providing a comprehensive view of complex specimens without the need for extensive sample preparation or the introduction of artifacts associated with deeper tissue penetration (Pampaloni et al., 2015; Vargas-Ordaz et al., 2021; Power and Huisken, 2017). These advantages have led light sheet microscopy to be used to probe 3D tissues at the cellular and subcellular levels (Reynaud et al., 2008) which minimizes user bias in evaluating the image (Buglak et al., 2021). Despite all the advantages of light sheet fluorescence microscopy, sample preparation can be challenging. Samples need to be transparent enough for the light sheet to penetrate, which often necessitates special clearing techniques that can be time-consuming and may alter the sample's natural state (see Section 4) (Delage et al., 2023).

## 2.5 Stimulated emission depletion microscopy (STED)

Stimulated Emission Depletion (STED) microscopy is a super-resolution imaging technique that overcomes the diffraction limit traditionally associated with optical microscopy (Calovi et al., 2021; Revelo et al., 2015). Stimulated Emission Depletion microscopy (STED) involves two lasers: an excitation laser that induces fluorescence in the targeted fluorophores, and a stimulated emission depletion (STED) laser that precisely quenches this fluorescence. The stimulated emission depletion (STED) beam is typically shaped like a doughnut, with a zero-intensity center coinciding with the focal point of the excitation beam pulsing continuously (Blom and Widengren, 2014; Wildanger et al., 2009; Willig et al., 2007; Wildanger et al., 2008). This configuration ensures that fluorescence is selectively depleted in the periphery but not at the center, thereby shrinking the effective area of light emission and enhancing the spatial resolution beyond the 200–300 nm (close to 25 nm) limit typical of conventional light microscopes (Figure 3D) (Angibaud et al., 2020; Müller et al., 2012; Blom and Brismar, 2014; Vicidomini et al., 2018; Harke et al., 2008). Optimizing STED microscopy involved focusing on precise temporal alignment between excitation and STED pulses and adjusting the polarization state of the STED beam. Time-gated detection was used to compensate for timing difference, and maintaining circular polarization of the STED beam to ensure effective fluorescence quenching at the zero-intensity point, which significantly enhanced image resolution and clarity (Galiani et al., 2012). However, continuous-wave pulses in STED microscopy can lead to increased photodamage, photobleaching, and phototoxic effects due to prolonged exposure of biological samples to intense light.

# 3 Exploiting photo-physics of materials in microscopy

In the prior section, we discussed imaging innovations which center on controlling sample illumination and emitted photon detection. A second, parallel means for enhancing spatiotemporal resolution exploits an ever-growing body of knowledge of photo-physics, which details how materials respond to light. Changes in the local environment of an electron can increase the likelihood of the photon-absorption-triggered excited state to decay by non-radiative means (e.g., through pathways besides emission of a lower-energy photon). Because material characteristics (e.g., chemical composition) can be selected to control electrons' environment in a predictable manner, it has been possible to engineer key fluorescence properties of fluorophores (e.g., quantum yield, Stokes shift) (Lakowicz and Masters, 2008). Here, we detail several imaging modalities that exploit our growing knowledge of photo-physics to explore subcellular structures. A persistent theme in this body of work is that photophysical properties of materials can be exploited to overcome the diffraction limit in resolving nanoscale structures.

## 3.1 FRET, FLIM, FRAP microscopy

Förster resonance energy transfer (FRET), Fluorescence Lifetime Imaging Microscopy (FLIM), and Fluorescence Recovery After Photobleaching (FRAP) represent sophisticated fluorescence microscopy techniques that each play a unique role in elucidating molecular dynamics within live cells by leveraging the photophysical properties of fluorescence molecules to surpass the diffraction limit. These techniques not only provide insights into the spatial and temporal aspects of molecular interactions but also offer a window into the dynamic cellular environments (Durhan et al., 2023; Yasuda, 2012; Kong et al., 2007; Huebsch and Mooney, 2007).

### 3.1.1 Förster resonance energy transfer (FRET)

FRET is particularly powerful for studying molecular interactions at the nanoscale. This technique is a non-radiative process which relies on the energy transfer between closely spaced donor and acceptor fluorophores, typically within 1–10 nm by dipole-dipole interactions. This energy transfer occurs when the donor's emission spectrum overlaps with the acceptor's absorption spectrum. Essentially, the donor fluorophore becomes excited, and its emission stimulates the acceptor, which then emits fluorescence. This allows for the detection of interactions between closely packed molecules (Figure 4A) (Gordon et al., 1998). Resolving interactions over such a small distance is challenging with optical approaches, including super-resolution microscopy. FRET efficiency, which reflects molecular proximity, is highly sensitive to the distance between these fluorophores, making it an excellent tool for detecting and quantifying protein-protein interactions, enzyme activities, and conformational changes within cells with high spatial and temporal resolution. This allows precise mapping of molecular interactions, and supports real-time monitoring of dynamic molecular events, as energy transfer between fluorophores occurs almost instantaneously, this facilitates the observation of processes (Coelho et al., 2020; Badia-Soteras et al., 2020; Broussard and Green, 2017; Padilla-Parra and Tramier, 2012; Day and Davidson, 2012). Recent advancements in FRET technology have been driven by the development of enhanced fluorescent proteins and sensitive microscopy equipment that enable the detailed visualization of molecular interactions and the real-time tracking of intracellular processes, such as signal transduction (Pietraszewska-Bogiel and Gadella, 2011). Intensity-based methods for detecting FRET are relatively straightforward and can be implemented with standard wide-field or confocal microscopes, using ratiometric FRET sensors to simplify data acquisition and analysis. However, interpreting changes in fluorescence intensity in FRET experiments can be complex when donor and acceptor fluorophores are not uniformly distributed, and are susceptible to artifacts from photobleaching. Fluorescence Lifetime Imaging Microscopy (FLIM) offers a robust alternative (Pietraszewska-Bogiel and Gadella, 2011).

### 3.1.2 Fluorescence lifetime imaging microscopy (FLIM)

In FLIM, a fluorophore is excited by laser, and the time it takes for the emitted photon to return to the ground state is recorded. FLIM complements FRET by providing the temporal dimension of fluorescence, measuring the decay time of the fluorescence emission from fluorophores (Mannam et al., 2021; Llères et al., 2017). Unlike intensity-based imaging, FLIM is not affected by changes in fluorophore concentration or local variations in brightness, which makes it suitable for accurate quantification of FRET efficiency (Elson et al., 2004). This attribute of FLIM is valuable in live-cell imaging, where it is used to explore changes in the molecular environment that influence fluorescence lifetime, such as pH variations or ion concentrations (Wang et al., 2019). FLIM's ability to provide a direct measure of the molecular environment enhances its application in monitoring the functional status of proteins and other biomolecules, facilitating a deeper understanding of cellular metabolism and disease pathologies (Wallrabe and Periasamy, 2005). The integration of FRET and FLIM techniques leverages FLIM's ability to measure changes in the fluorescence lifetime of a donor molecule upon energy transfer to an acceptor, independent of fluorophore concentration or excitation intensity which is useful in quantifying molecular proximity and interactions within complex biological systems (Llères et al., 2009), without the limitations seen in intensity-based methods providing higher spatial and temporal (nanometer and nanosecond) resolution (Coelho et al., 2020; Padilla-Parra and Tramier, 2012; Wallrabe and Periasamy, 2005; Elangovan et al., 2002; Poland et al., 2015). However, FLIM requires specialized equipment and complex data analysis, and may need longer integration times, potentially limiting its application in rapidly changing processes (Pietraszewska-Bogiel and Gadella, 2011).

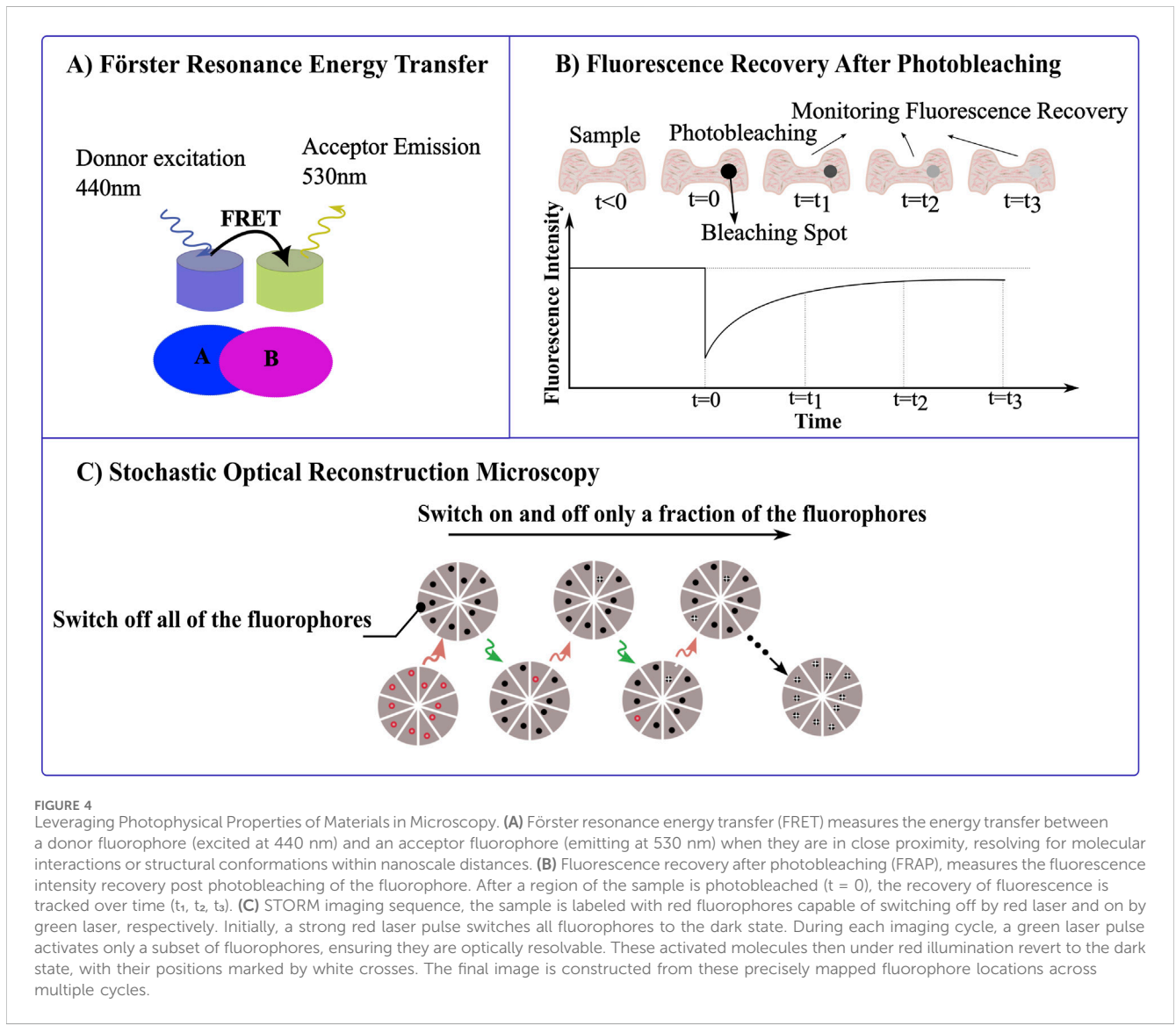

FIGURE 4
Leveraging Photophysical Properties of Materials in Microscopy. **(A)** Förster resonance energy transfer (FRET) measures the energy transfer between a donor fluorophore (excited at 440 nm) and an acceptor fluorophore (emitting at 530 nm) when they are in close proximity, resolving for molecular interactions or structural conformations within nanoscale distances. **(B)** Fluorescence recovery after photobleaching (FRAP), measures the fluorescence intensity recovery post photobleaching of the fluorophore. After a region of the sample is photobleached (t = 0), the recovery of fluorescence is tracked over time ($t_1$, $t_2$, $t_3$). **(C)** STORM imaging sequence, the sample is labeled with red fluorophores capable of switching off by red laser and on by green laser, respectively. Initially, a strong red laser pulse switches all fluorophores to the dark state. During each imaging cycle, a green laser pulse activates only a subset of fluorophores, ensuring they are optically resolvable. These activated molecules then under red illumination revert to the dark state, with their positions marked by white crosses. The final image is constructed from these precisely mapped fluorophore locations across multiple cycles.

### 3.1.3 Fluorescence recovery after photobleaching (FRAP)

Fluorescence recovery after photobleaching (FRAP) is a technique that is used to study the dynamics of fluorescence probes by analyzing the recovery of fluorescence in a photobleached area over time. In this technique, a specific region of fluorescent molecules is photobleached using high-intensity light. Over time, fluorescence gradually reappears as unbleached molecules from surrounding areas diffuse into the bleached zone. Fluorescence recovery of the acceptor after photobleaching is a straightforward technique that does not require specialized equipment and can be used to overcome artifacts associated with intensity-based FRET analysis. FRET can be calculated by monitoring the increase in donor fluorescence after acceptor photobleaching. Moreover, this method is destructive and unsuitable for repeated observations in dynamic studies, as it relies on irreversible photobleaching of the acceptor (Ishikawa-Ankerhold et al., 2012; Swift and Trinkle-Mulcahy, 2024).

By bleaching a region of fluorescence and observing how quickly and completely the fluorescence returns, it can measure the kinetics of molecular movement and interaction within that region (Figure 4B). This technique is particularly useful for studying the dynamics of membrane proteins, cytoskeletal elements, and nucleic acids, which provides insights into cellular trafficking, membrane dynamics and chromatin remodeling processes (Swift and Trinkle-Mulcahy, 2024; De Los Santos et al., 2015).

### 3.2 Stochastic optical reconstruction microscopy (STORM)

Stochastic Optical Reconstruction Microscopy (STORM) is a super-resolution imaging technique that surpasses the traditional diffraction limits of light microscopy (Xu et al., 2017; Rust et al., 2006; Veeraraghavan et al., 2016). It is predicated on the principle of precise localization of individual fluorescent molecules that are activated stochastically. In STORM, specific fluorophores capable of photo-switching are employed; these fluorophores can alternate between bright (fluorescent) and dark (non-fluorescent) states when exposed to light of particular wavelengths (Pfender et al., 2014;

Hainsworth et al., 2018; Wu et al., 2013). The fundamental physics of STORM relies on the ability to accurately detect and localize the positions of these sparsely illuminated fluorophores during each imaging cycle (Figure 4C) (Codron et al., 2021). The fact that only small fraction of fluorophores is able to emit light at any given time, means that their separation exceeds the diffraction limit. The resolution achievable with STORM, which can go down to about 20 nm lateral and 50 nm axial, is considerably enhanced by the photophysical properties of the dyes used, such as their ability to emit sufficient photons for accurate localization and their capacity for multiple, controlled switching cycles without significant photobleaching (Hainsworth et al., 2018; Samanta et al., 2019; Tam and Merino, 2015; Van De Linde et al., 2011; Xia and Fu, 2024). Compared to other super-resolution techniques like STED, STORM provides higher spatial resolution and does not require high-intensity or continuous illumination of lasers, which can cause photodamage to the samples (Tam and Merino, 2015). This makes STORM particularly advantageous for detailed, high-contrast imaging of molecular assemblies within cells with minimal sample damage. It has been expected that STORM will prove to be a useful instrument for immunofluorescence imaging and high-resolution fluorescence *in situ* hybridization (Rust et al., 2006).

## 3.3 Reflectance confocal microscopy

Confocal microscopy can be performed in reflectance mode, without the use of fluorescent dyes, to provide detailed images of tissue architecture and cellular morphology of living tissue in near real-time (Clark et al., 2003). This method of confocal imaging with reflected light relies on the differential backscattering properties from cellular morphology and tissue architecture to provide contrast. Essentially, in this method, a focused laser beam is used to illuminate the tissue, and light that is backscattered from different part of the tissue structure is collected. A pinhole is placed in front of the detector, allowing only the light from the exact focal plane to be captured, while out-of-focus light is blocked (Liang et al., 2009). Hence, it resembles histological tissue evaluation, except that the subcellular resolution is achieved noninvasively and without stains or dyes (Rudnicka et al., 2008). Although this method is promising for real-time longitudinal studies, it still has penetration limitations, particularly for highly scattering tissues (Tan et al., 2004; Liang et al., 2009; Rudnicka et al., 2008).

## 3.4 Second harmonic generation (SHG)

SHG, a nonlinear optical process in which two photons with the same frequency interact with a specimen and are effectively combined to form a new photon with twice the energy and therefore twice the frequency and half the wavelength of the original photons. This method facilitates the imaging of non-centrosymmetric structures like collagen fibers without external labeling, providing a clear view of the structural organization and integrity of biological tissues. This capability is especially beneficial for studying connective tissues and muscle fibers, offering a non-invasive tool to explore the intricate architectures of various structural proteins within their native environments (Ustione and Piston, 2011; Zipfel et al., 2003; Lefort, 2017).

## 3.5 Two-photon autofluorescence

Two-Photon Autofluorescence leverages the natural fluorescence of biomolecules such as Nicotinamide Adenine Dinucleotide (NADH) and flavoproteins to probe cellular metabolism to provide comprehensive insights into tissue composition and functionality. When it is combined with Fluorescence Lifetime Imaging Microscopy, this technique can further analyze metabolic states, revealing critical details about tissue health, disease progression, and potential therapeutic effects (Neto et al., 2020).

## 3.6 Third harmonic generation

Third Harmonic Generation (THG), where three photons with the same frequency interact with a specimen and are effectively combined to form a new photon with thrice the energy, occurs at interfaces with significant refractive index changes, such as cellular membranes and lipid boundaries, which provides detailed imaging of tissue interfaces and morphological boundaries without the need for dyes (Schenke-Layland et al., 2005; Borile et al., 2021; Hoover and Squier, 2013; Lilledahl et al., 2007; Schenke-Layland, 2008).

# 4 Optimizing sample processing for improved resolution in 3D engineered tissue imaging

In the prior section, we discussed about enhancing spatiotemporal resolution through the exploitation of photophysical properties of materials, which details how materials respond to light. A third method to enhance imaging resolution involves optimizing sample processing. The way a sample is processed can directly affect the image resolution by understanding how light interacts with the tissue. Refining aspects like sample size, transparency, and fixation methods, we enhance light penetration and minimize light scattering. This integration of advance material processing with optical techniques opens a new area for capturing detailed cellular structures with high clarity. Here, we will elaborate on various methods that have been employed to optimize sample processing for better imaging compatibility.

## 4.1 Tissue engineering miniaturization techniques for enhanced sample imaging compatibility

Miniaturization of engineered tissues significantly enhances their compatibility with advanced imaging techniques, enabling high-resolution, real-time visualizations which is crucial for detailed studies of cellular dynamics (Bose et al., 2019; Bhatia and Ingber, 2014). In organs-on-chips, precise positioning of cell types relative to each other simplifies the integration of fluorescence microscopy (Bose et al., 2019; Kutys et al., 2020). Reducing tissue size minimizes light scattering, which is a significant challenge in imaging larger tissues. Smaller tissues ensure that light can penetrate more deeply and evenly, which enhances the clarity and detail of images (Wang et al., 2003). Techniques such as microfabrication, microfluidics, and 3D bioprinting including photoencapsulating cells in polyethylene glycol hydrogels (Chen et al., 2010), two-step SU-8 lithography (Boudou et al., 2012), two photon direct laser writing (TPDLW) (Michas et al., 2022), hydrogel-assisted double molding (Simmons et al., 2023) are employed to create tissues that accurately mimic organ-level functions on a microscale (Michas et al., 2022; Huebsch et al., 2016). Also, these micro-engineered tissues facilitate high-throughput drug screenings and disease modeling by allowing parallelized assays (Thavandiran et al., 2013).

## 4.2 Advanced fixation strategies for 3D engineered tissues

Fixation of 3D engineered tissues is essential for maintaining their structural integrity and molecular composition, and also critical for accurate histological evaluation and molecular analysis (Howat and Wilson, 2014; Sampedro-Carrillo and Del Valle, 2022; Agar et al., 2007; Da Silva et al., 2023). Traditional formalin-based approaches like neutral buffered 4% formaldehyde with added 2% phenol which form cross-links between proteins, have proven effective in preserving various tissue morphology, but can modify proteins and nucleic acids, affecting molecular assays (Howat and Wilson, 2014; Sampedro-Carrillo and Del Valle, 2022; Sabatini et al., 1963; Groelz et al., 2013; Rodgers et al., 2021; Buesa, 2008; Sánchez-Porras et al., 2023; Hopwood et al., 1989; Serrato et al., 2024; Eltoum et al., 2001; Grizzle, 2009). Alcohol-based fixatives, while better at preserving molecular integrity for such assays, may not provide the same level of morphological detail as formalin (Howat and Wilson, 2014; Gillespie et al., 2002; Kap et al., 2011; Lussier et al., 2023). Among these fixatives, the PAXgene System offers a non-toxic method that combines alcohols, acetic acid, and soluble organic compounds, effectively preserving tissue morphology while ensuring biochemical stability which makes it ideal for transportation and long-term preservation (Groelz et al., 2013; Lahiri et al., 2021). For lipid-rich tissues, osmium tetroxide stabilizes lipids without compromising their integrity (Sánchez-Porras et al., 2023). Recent advancements include molecular fixatives like Hepes-glutamic acid buffer mediated Organic solvent Protection Effect (HOPE), which uses a Hepes-glutamic acid buffer and acetone, and Universal Molecular FIXative (UMFIX), a methanol-based fixative with polyethylene glycol, both designed to preserve molecular integrity better than formaldehyde solutions (Howat and Wilson, 2014; Shuster et al., 2011; Koch et al., 2012). Periodate-lysine-paraformaldehyde (PLP) method enhances fixation by adding periodate that oxidizes polysaccharides, forming additional cross-links, and Bouin's fixative is excellent for delicate tissues, preserving fine cellular details and specific structures like glycogen and nuclei, though it has slow penetration and can distort certain tissues (Hewitson et al., 2010). These fixatives provide a balance between preserving tissue morphology and facilitating molecular analysis, crucial for achieving accurate and reliable results in diagnostic and research settings, especially when dealing with complex 3D tissue structures. This balance makes them particularly suitable for light imaging applications where detailed morphological examination is as crucial as molecular integrity.

## 4.3 Tissue clearing techniques to enhance transparency

Tissue clearing is a post-processing method that removes lipids from biological tissue, and transforms opaque tissues into transparent samples for detailed three-dimensional imaging within intact tissues (Yu et al., 2021; Ueda et al., 2020; Tomer et al., 2014; Mai and Lu, 2024). Various methods, including organic solvent-based, aqueous-based, and hydrogel embedding techniques, each come with specific advantages and drawbacks. Organic solvent-based methods such as 3DISCO (3D Imaging of Solvent-Cleared Organs), BABB (Benzyl Alcohol and Benzyl Benzoate), and DBE (Dibenzyl Ether) are fast and effective but may diminish fluorescent signals and alter structural integrity using organic solvent (Ueda et al., 2020; Mai and Lu, 2024; Kolesová et al., 2021; Jing et al., 2019; Lu et al., 2022; Tian et al., 2021; Muntifering et al., 2018; Brenna et al., 2022). Aqueous-based methods like CUBIC (Clear, Unobstructed Brain/Body Imaging Cocktails) and SeeDB (See DEEP Brain) use hydrophilic substances and sugars (e.g., fructose) to gently enhance transparency while preserving fluorescence while reducing the tissue autofluorescence, although they may not achieve complete transparency (Ueda et al., 2020; Mai and Lu, 2024; Kolesová et al., 2021; Jing et al., 2019; Tian et al., 2021; Muntifering et al., 2018; Brenna et al., 2022). Hydrogel embedding techniques, including CLARITY (Clear Lipid-exchanged Acrylamide-hybridized Rigid Imaging/Immunostaining/*In situ* hybridization-compatible Tissue-hYdrogel), PACT (PAsive Clarity Techniques), PARS (Perfusion-assisted Agent Release *in Situ*), and SCALE (Sorbitol Clearing Agent for Light-sheet microscopy Enhancement) maintain molecular and structural integrity through a hydrogel-tissue hybrid to support extensive molecular labeling. These approaches integrates well with advanced imaging technologies like light-sheet fluorescence microscopy, and are ideal for high-resolution imaging, despite their complexity and the longer time required for processing (Ueda et al., 2020; Tomer et al., 2014; Mai and Lu, 2024; Kolesová et al., 2021; Jing et al., 2019; Lu et al., 2022; Tian et al., 2021; Muntifering et al., 2018; Brenna et al., 2022; Costantini et al., 2019; Du et al., 2018; Gómez-Gaviro et al., 2020).

## 4.4 Expansion microscopy

Expansion microscopy (ExM) is an innovative imaging technique that enhances resolution by physically enlarging fixed biological samples that are embedded in a swellable hydrogel to achieve <120 nm expansion-corrected lateral resolution (Günay et al., 2023). The fundamental principle behind expansion microscopy involves infusing the fixed biological samples with a monomer solution, which is then polymerized into a hydrogel. As water is introduced, the hydrogel swells, causing the sample to expand isotropically while maintaining the relative position and structure of the embedded sample (Chen et al.). This technique has been developed to overcome the limitations of traditional and super-resolution microscopy, which require expensive, specialized equipment (Chozinski et al., 2016). Expansion microscopy uses various types of hydrogels, such as polyacrylate, polyethylene glycol (Gao et al., 2021) and a more advanced formulation of N,N-dimethylacrylamide and sodium acrylate (Truckenbrodt et al., 2018), to achieve different expansion factors and resolutions. Initial methods achieved approximately fourfold expansion, while newer approaches like the 10X method allow for tenfold expansion, achieving resolutions down to 25 nm on standard fluorescence microscopes by just changing the gel formulation (Günay et al., 2023; Truckenbrodt et al., 2018; Tillberg et al., 2016; Zhao et al., 2017). Furthermore, expansion microscopy has been adapted to specifically target and visualize RNA through the development of "ExFISH" (Expansion Fluorescence *In Situ* Hybridization). Which incorporates a small-molecule linker to covalently anchor RNA to a swellable polyelectrolyte gel used in expansion microscopy. This method allows for the detailed examination of RNA within cells and intact tissue, crucial for understanding its role in gene expression and cellular function at nanoscale resolution within tissues (Chen et al., 2016).

A significant expansion microscopy challenge is its dependence on a costume-made probes for each type of protein to make sure that the fluorescent labels are accurately attached and positioned relative to the protein target throughout the gel expansion, which is costly and time-consuming to produce. To address this challenge, a method to use conventional, commercially available fluorophore-labeled antibodies and intrinsic fluorescent proteins directly in expansion microscopy have been developed, by introducing new chemical linking strategies that retain these labels during and after the expansion process, such as small, amine-reactive molecules like methacrylic acid N-hydroxysuccinimidyl ester (MA-NHS) and glutaraldehyde (GA). These molecules effectively link the antibodies and fluorescent proteins to the expandable hydrogel matrix, which ensures they remain in place and maintain their fluorescence after the physical expansion of the sample (Chozinski et al., 2016). A disadvantage of this approach is the inability to re-probe the expanded sample for different antigens.

# 5 Exploiting advanced optics, fluorescent probes and sample preparation to reveal subcellular architectures in tissue engineering

Building upon the advancements in imaging technology, the combination of sophisticated optical techniques, fluorescent probes, and sample preparation techniques has significantly enhanced our ability to explore and visualize subcellular structures in tissue engineering. For instance, Spinning Disc Confocal Microscopy have used a commercially fabricated chromium photomask to enable 3D biological imaging across a wide wavelength range (400–800 nm). This innovation incorporates hundreds to thousands of simultaneous illumination points, reducing peak illumination power density (Halpern et al., 2022). Furthermore, integration of fluorescence lifetime imaging microscopy by an Spinning Disc Confocal Microscopy into a frequency-domain FLIM system, reduces out-of-focus blur and photobleaching while maintaining short acquisition times (Van Munster et al., 2007). In order to get high temporal resolution to study calcium signaling in vascular endothelial cells, Spinning Disc Confocal Microscopy equipped with an electron-multiplying CCD camera has been used to reveal rapid and localized calcium transients which is critical in regulating blood flow and pressure (Nelson et al., 2012). Furthermore, optical photon reassignment (OPR), an advanced super-resolution method using Spinning Disc Confocal Microscopy, have been used to significantly enhance lateral resolution by a factor of 1.37 through a single exposure (Azuma and Kei, 2015). Moreover, to achieve high resolution and contrast in live 3D engineered tissues, a custom-built Spinning Disc Confocal Microscopy with the second near-infrared (NIR-II) optical window have been used to achieve lateral (0.5 ± 0.1 μm) and axial (0.6 ± 0.1 μm) resolution (Zubkovs et al., 2018). Lastly, the combination of engineered tissues with spinning disc confocal microscopy has given us new insights into mitochondrial dynamics. Using Spinning Disc Confocal Microscopy, Ahmadian et al. optimized and quantified live images of 3D mitochondrial networks in mesoangioblasts and mesoangioblast-derived myotubes. This technique reduced background signals and variation in fluorescence intensity, enabling more accurate and reproducible imaging compared to Laser Scanning Confocal Microscopy. This allowed for the identification of key changes in mitochondrial structure and function during cell differentiation, which enhances our understanding of mitochondrial behavior in complex tissue environments (Ahmadian et al., 2024).

Multiple light-sheet microscopy enhances the capabilities of single plane illumination microscopy by reducing photobleaching and enhancing imaging speed and is particularly suitable for imaging. This approach uses spatial filters in the excitation arm to create multiple light-sheets to allow simultaneous illumination of multiple planes within a sample, to facilitate rapid imaging. The multiple light-sheet microscopy system demonstrated the ability to produce thinner light-sheets with increased resolution and reduced crosstalk between the illuminated planes (Mohan et al., 2014). Light-sheet microscopy techniques enhanced by reversible saturable optical fluorescence transitions (RESOLFT) significantly surpass the diffraction limit in axial resolution, enabling 3D imaging of live biological specimens with minimized photodamage and light exposure and it has the ability to produce optical sections up to 5–12 times thinner than those achievable with conventional diffraction limited light-sheet fluorescence microscopy (Hoyer et al., 2016). Advancements in light-sheet fluorescence microscopy have significantly improved high-speed 3D imaging of live tissues, especially for calcium dynamics in cardiac engineered tissues. Light-sheet fluorescence microscopy offers enhanced imaging speed and resolution with minimal light exposure, providing deeper insights into fast physiological processes. Sparks et al. used LSFM to obtain high-resolution images of t-tubule networks in real tissues, revealing their precise 3D structure and organization, which is critical for efficient excitation-contraction coupling. This level of detail is challenging to achieve with laser scanning confocal or two-photon microscopy due to their limitations in depth penetration and resolution (Sparks et al., 2020).

Two-photon microscopy has been used to study the enhancement of the structural and functional properties of engineered heart tissues through chronic electrical stimulation, which is crucial to examine the maturation and functionality of tissue-engineered cardiac constructs. In the study, two-photon microscopy revealed detailed visualization of increased connexin-43 abundance, indicating better gap junction formation and electrical coupling in paced engineered heart tissues, and enabled the measurement of a significantly thicker compact cardiomyocyte layer in paced engineered heart tissue and demonstrated a more homogeneous distribution of cardiomyocytes throughout the tissue depth (Hirt et al., 2014). Two-photon microscopies with near-infrared femtosecond laser pulses have been used to visualize the intratissue elastic fibers within both native and engineered heart valves without the need for invasive procedures such as tissue removal, embedding, or staining. Using multiphoton-induced autofluorescence and second harmonic generation, König et al. could clearly differentiate between elastic fibers and collagenous structures within the extracellular matrix (König et al., 2005). Blazeski et al. used two photon microscopies to visualize and analyze cellular and subcellular structures in engineered heart slices (EHS). This allowed for the precise localization of live cells relative to the matrix and the observation of structural cues promoting cellular alignment, which are crucial for understanding the physiological and pathophysiological roles of the ECM in cardiac tissue engineering (Blazeski et al., 2015). Moreover, to demonstrate that intermittent mechanical straining enhances and accelerates collagen fiber alignment in engineered heart tissues, two. photon microscopy have been used to achieve high resolution in thick 3D engineered tissues (Rubbens et al., 2009). Ye et al. used two-photon laser scanning microscopy to show that SDS best preserves the structural integrity of the 3D engineered heart tissues as decellularization reagent and visualize details about the microstructure of decellularized porcine heart tissue. With this technique, they were able to observe the ultrastructural morphology of type I collagen within the extracellular matrix (ECM) without the need for complicated fixation and washing processes, which can alter the tissue's natural structure (Ye et al., 2016).

Expanding beyond cardiovascular applications, Chang et al. used third harmonic generation (THG) and two-photon excited fluorescence (2PEF) imaging techniques to non-invasively monitor the functional properties of 3D engineered human adipose tissues to provide detailed, label-free images of lipid droplet formation and cell metabolism over time. THG was used to visualize lipid droplets, while 2PEF assessed the metabolic state of cells through the redox ratio of endogenous FAD and NADH fluorescence. The combination of these imaging approach allowed this team to dynamically correlate lipid accumulation with changes in the metabolic state during adipogenic differentiation. They discovered that cells in adipogenic media showed a significantly lower redox ratio and increased lipid content compared to those in propagation media, providing new insights into the relationship between metabolic changes and lipid biosynthesis in differentiating stem cells (Chang et al., 2013). Meanwhile, Straub et al. used Multifocal Multiphoton Microscopy to achieve non-invasive, high-resolution live cell imaging, which significantly enhanced the understanding of cell dynamics and morphology in three dimensions. This method used an array of high numerical aperture foci for parallel multiphoton excitation, allowing rapid and detailed 3D reconstructions of live cells at speeds up to video rate. This approach is particularly advantageous for dynamic cellular processes that require fast imaging techniques. For instance, they were able to visualize the detailed morphology of neurites and the dynamics of membrane-associated processes, which were not as comprehensively understood with previous imaging technologies due to slower acquisition times and greater photodamage (Straub et al., 2000). To image 3D engineered neuronal networks, a hybrid multiphoton microscope combining scanning-line temporal-focusing with laser-scanning two-photon microscopy have been introduced. This dual approach allows for both rapid volumetric imaging and detailed structural analysis within a single system, which is crucial for investigating dynamic neuronal activities and interactions in 3D cultures. The primary achievement of this method is highly detailed three-dimensional imaging at a microscopic scale across a volume of tissue and at high speeds (tens of volumes per second). This has enabled the observation of over 1,000 developing cells and their complex spontaneous activity patterns within millimeter-scale structures. This system use a regeneratively amplified ultrafast laser to enhance two-photon absorption significantly, which improves the signal-to-noise ratio (Dana et al., 2014).

Super-resolution Stimulated Emission Depletion (STED) microscopy has been used for 2D and 3D cell culture methods to study the molecular organization of apical and lateral membrane domains of epithelial cells. By using an inverted filter mounting strategy, Maraspini et al. were able to gain better access to the apical membrane domains, allowing them to effectively image and resolve the densely packed micro-villi of human enterocytes. Additionally, the optimization of 3D organotypic cell culture enabled the detailed visualization of adhesion complexes in the lateral membrane domain of kidney-derived cells which significantly contributed to our understanding of epithelial membrane organization (Maraspini et al., 2020). Moreover, STED microscopy was utilized to visualize dynamic morphological changes in dendritic spines with a resolution of about 70 nm, significantly enhancing the understanding of synaptic function and plasticity beyond what is possible with confocal microscopy (Nägerl et al., 2008), and STED nanoscopy was used for detailed visualization of mitochondrial dynamics and interactions at high resolution (Liu et al., 2022).

Multicolor three-dimensional stochastic optical reconstruction microscopy (3D STORM) has been used to examine cellular structures and their interactions at nanoscale resolution. By this method Huang et al. were able to image entire mitochondrial networks in mammalian cells, which shows detailed mitochondrial morphologies and their spatial relationships with microtubules that were not visible through epifluorescence microscopy (Huang et al., 2008b). Moreover, STORM was used to examine the nanoscale distribution and clustering characteristics of Angiotensin II Receptor Type 1 (AT1R) on the membrane of PC12 cells. Aldossary et al. found that exposure to hypoxia for 24 h increased the maximum cluster area, suggesting a formation of superclusters. This change in the clustering behavior of AT1Rs under hypoxia might influence the receptor's function and cell signaling pathways, indicating a potential mechanism by which cells respond to low oxygen levels (Aldossary et al., 2023). Lastly, STORM was used to visualize the endothelial surface glycocalyx (ESG) of cultured endothelial cells, revealing its molecular components and ultrastructure with high spatial resolution (Xia and Fu, 2024).

## 6 Conclusion

In this review, we have highlighted the growing importance of high-resolution sub-cellular imaging in the field of tissue engineering. These complex structures offer a more realistic environment for studying cellular behavior and disease compared to traditional 2D cultures. However, obtaining high-resolution images of these dense tissues presents a challenge due to limitations in light penetration and accessibility. Fortunately, advancements in microscopy techniques (like light sheet fluorescence microscopy), fluorescent probes, and material processing are paving the way for us to overcome these limitations. This allows for detailed visualization of sub-cellular structures within 3D tissues, providing valuable insights into critical cellular processes and their response to engineered environments.

We have also discussed the importance of balancing various factors like spatial and temporal resolution, signal-to-noise ratio, and photobleaching to achieve optimal image quality. This review serves as a comprehensive guide for researchers navigating the growing toolbox of imaging techniques available for studying 3D engineered tissues and organoids, ultimately leading to a deeper understanding of cell-cell interactions, disease mechanisms, and the development of improved *in vitro* models for drug discovery and therapy.

## 7 Future work

High-resolution sub-cellular imaging in tissue engineering has a bright future ahead of it. For the purpose of investigating larger and more complex 3D tissues, it will be important to continue developing improved microscopy techniques with even deeper tissue penetration and decreased phototoxicity. One of these emerging methods is Single-Molecule Orientation-Localization Microscopy that uses fluorogenic probes that specifically bind to the structure of interest; thus, it detects and analyzes signals from individual molecules to achieve higher resolution by capturing detailed orientation and positional data (Zhou et al., 2024). Additionally, Single-Molecule Orientation-Localization can use a radially and azimuthally polarized multi-view reflector microscope to enhance the precision of measuring 6D spatial and orientational coordinates (3D position and 3D orientation) of single molecules within engineered tissues (Zhang et al., 2023). Lattice Light Sheet Microscopy is another emerging optical technique which uses a rapidly moving thin plane of light sheet. Unlike light-sheet fluorescence microscopy, in which the light sheets are uniform, the light sheet here is structured into a lattice pattern (Daugird et al., 2024). Also, the investigation of fluorescent probes to improve luminosity, biocompatibility, and multi-color capabilities will enable the concurrent imaging of several cellular structures and functions using these optical methods. Lastly, in order to process the massive amounts of data produced by these high-resolution imaging techniques efficiently, the development of automated image analysis techniques will be necessary. Likewise, approaches to glean useful information from these massive data sets, including machine learning, will be crucial (Kalkunte et al., 2024).

## Author contributions

YK: Conceptualization, Visualization, Writing–original draft, Writing–review and editing. NH: Conceptualization, Funding acquisition, Supervision, Writing–original draft, Writing–review and editing.

## Funding

The author(s) declare that financial support was received for the research, authorship, and/or publication of this article. YK and NH were supported by NIH grants R01HL107594, RO1HL159094, and NSF Grant CAREER 2338931.

## Acknowledgments

We thank Ganesh Malayath for his input in conceptualizing the figures.

## Conflict of interest

The authors declare that the research was conducted in the absence of any commercial or financial relationships that could be construed as a potential conflict of interest.

## Publisher's note

All claims expressed in this article are solely those of the authors and do not necessarily represent those of their affiliated organizations, or those of the publisher, the editors and the reviewers. Any product that may be evaluated in this article, or claim that may be made by its manufacturer, is not guaranteed or endorsed by the publisher.